\documentclass[%
reprint,
superscriptaddress,
unsortedaddress,
 amsmath,amssymb,
 aps,
pra,
floatfix,
]{revtex4-1}

\usepackage{graphicx}
\usepackage{dcolumn}
\usepackage{bm}

\begin{document}

\title{The DarkLight Experiment: A Precision Search for New Physics at Low Energies}
\thanks{Spokesmen: P. Fisher and R. Milner}\author{J. Balewski, J. Bernauer, J. Bessuille, R. Corliss, R. Cowan, C. Epstein, P. Fisher, D. Hasell, 
E. Ihloff,  Y. Kahn, J. Kelsey,  R. Milner, S. Steadman, J. Thaler, C. Tschal\" ar, C. Vidal} 
\affiliation{Laboratory for Nuclear Science and Bates Linear Accelerator Center, Massachusetts Institute of Technology, Cambridge, MA 02139} 
\author{S. Benson, J. Boyce, D. Douglas, P. Evtushenko, C. Hernandez-Garcia, C. Keith,  C. Tennant, S. Zhang}
\affiliation{Thomas Jefferson National Accelerator Facility, Newport News, VA 23606} 
\author{R. Alarcon D. Blyth, R. Dipert, L. Ice, G. Randall}
\affiliation{Arizona State University, Tempe, AZ 85287}
\author{B. Dongwi, N. Kalantarians, M. Kohl, A. Liyanage, J. Nazeer}
\affiliation{Hampton University, Hampton, VA 23668}
\author{M. Gar\c con}
\affiliation{CEA Saclay, Service de Physique Nucl\'eaire, 91191 Gif-sur-Yvette, France}
\author{R. Cervantes, K. Dehmelt, A. Deshpande, N. Feege }
\affiliation{Stony Brook University, Stony Brook, NY 11790}
\author{B. Surrow}
\affiliation{Temple University, Philadelphia, PA 19122}

\collaboration{\bf The DarkLight Collaboration}


\date{\today}

\begin{abstract}
We describe the current status of the DarkLight experiment at Jefferson Laboratory.  DarkLight is motivated by the possibility that a dark
photon in the mass range 10 to 100 MeV/c$^2$ could couple the dark sector to the Standard Model.  DarkLight will  precisely measure electron
proton scattering using the 100 MeV electron beam of intensity 5 mA at the Jefferson Laboratory energy recovering linac incident on a windowless gas target of molecular hydrogen.  The complete final state including scattered electron, recoil proton, and $e^+ e^-$ pair will be detected.  A phase-I experiment has been funded and is 
expected to take data in the next eighteen months.  The complete phase-II experiment is under final design and could run within two years after phase-I is completed.
The DarkLight experiment drives development of new technology for beam, target, and detector and provides a new means to carry out electron
scattering experiments at low momentum transfers.
\end{abstract}

\maketitle

\section{Motivation}
The search for new physics beyond the Standard Model is a major activity of nuclear and particle physicists worldwide.
A simple extension of the Standard Model Lagrangian can be proposed~\cite{Hol86, Ark08} by the addition of a kinetic mixing operator 
term $\frac{\epsilon}{2}F^Y_{\mu \nu} F^{\prime \mu \nu}$.  New ``dark" Abelian forces can couple to the Standard Model
hypercharge via $F^{\prime}_{\mu \nu} = \partial_{[\mu}A^{\prime}_{\nu]}$ and $A^{\prime}$ is a new dark gauge field.
If the $A^{\prime}$ is massive, Standard Model matter acquires milli-charges proportional to $\epsilon$ under the $A^{\prime}$. 
Impetus for searches for a new massive, dark photon come from both interpretations of recent astrophysical observations~\cite{Acc14, Agu14} of positron excess
as well as the 3-4$\sigma$ discrepancy between the measured value, by BNL experiment E821, of the anomalous magnetic moment of the muon and the
Standard Model expectation~\cite{Blu13}. If this is a real discrepancy, then it can be explained by a dark photon of mass $\sim$10 to 500 MeV/c$^2$.
Further, dark photons have been invoked to understand observed rates of 511 keV photons at the center of our galaxy~\cite{Boe09} as well as in the context of understanding
the proton radius puzzle~\cite{Ber14a}. 

The $A^{\prime}$ might be produced in a number of ways: radiative production in electron scattering from a nucleus; $\pi^0$ decay to $e^+e^- \gamma$;  $\Upsilon(2S,3S) \rightarrow \gamma A^{\prime}$, followed by $A{^\prime} \rightarrow \mu^+ \mu^-$; and in $e^+e^- \rightarrow \gamma A^{\prime}$,
followed by $A{^\prime} \rightarrow e^+e^-,\  \mu^+ \mu^-$ .
Over the last several years, there has been an intensive worldwide effort to search for evidence of  the $A^{\prime}$.  Ingenious searches have been 
conducted~\cite{Bjo09, E137} using data from past experiments completed within the last three decades.  In addition, running experiments built for other scientific purposes have been used to search~\cite{Babar, KLOE, HADES, WASA, PHENIX, MAMI} for the $A^{\prime}$.  At the LHC, both ATLAS~\cite{ATLAS} and CMS are conducting searches for weakly coupled light gauge bosons. At Fermilab, the SeaQuest experiment will perform a search~\cite{SeaQuest}.  To date, a large area of the coupling $\alpha^\prime \equiv \epsilon^2 \alpha_{EM}$ vs. mass ($m_{A^\prime}$) space for $A^\prime$ decay to $e^+e^-$ has been excluded at the 2$\sigma$ level. No evidence of the $A^{\prime}$ has been found.  
Further searches must aim for higher precision by using dedicated experiments employing innovative experimental techniques.

\section{The DarkLight Experiment}

\begin{figure}
\centering\includegraphics[width=0.4\textwidth,height=0.3\textheight]{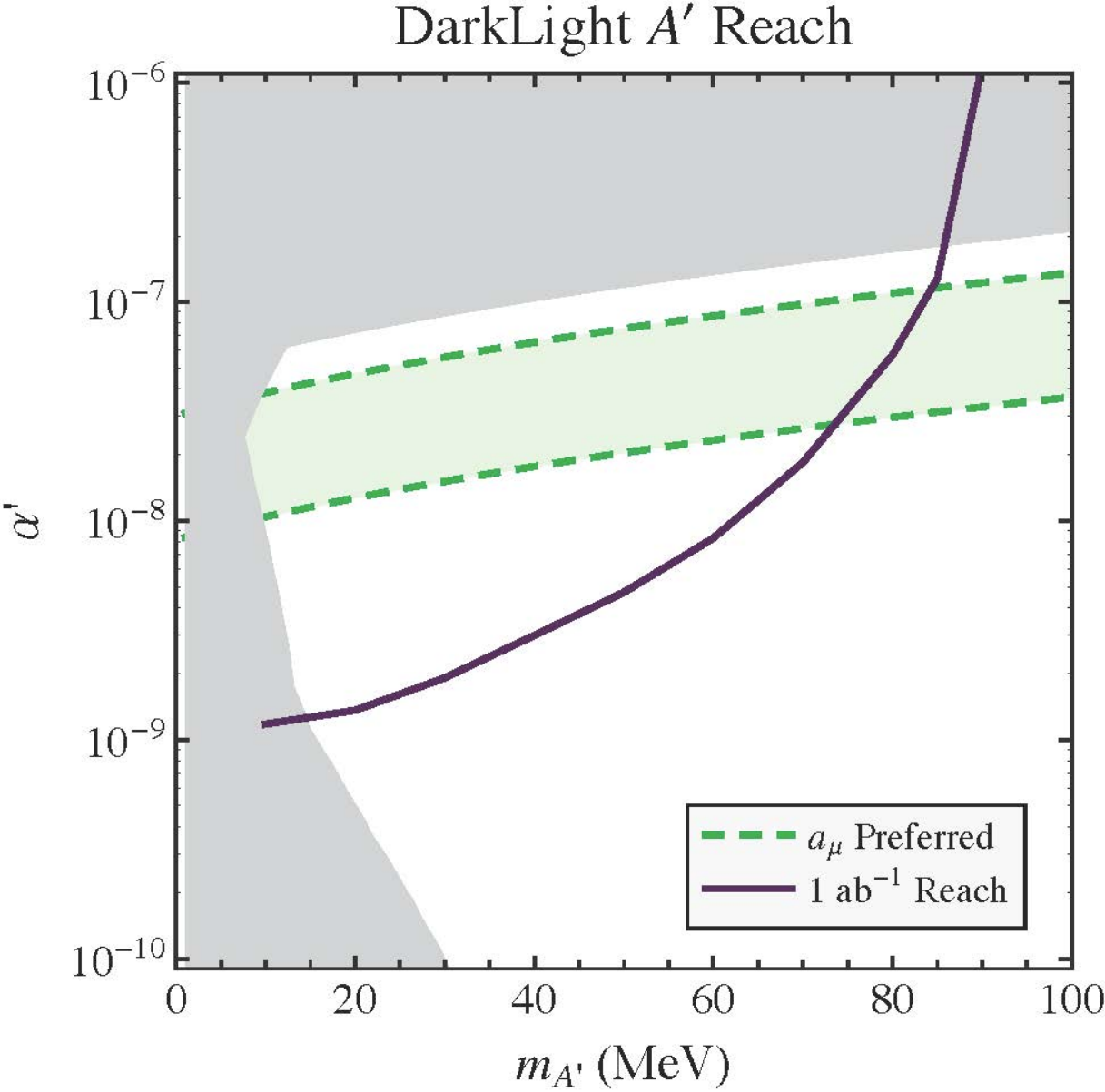}
\caption{The 5$\sigma$ limit reached by the complete DarkLight experiment in a run of 1 ab$^{-1}$ integrated luminosity~\cite{Thaler}  Searches with
negative results at the level of 2$\sigma$ cover the coupling-mass space proposed for DarkLight.  DarkLight is unique in searching for $A^\prime$ production
from the proton at low masses and will search for invisible decays, which are weakly excluded.}
\label{Sensitivity}
\end{figure}

\begin{figure*}
\centering\includegraphics[width=0.9\textwidth,height=0.4\textheight]{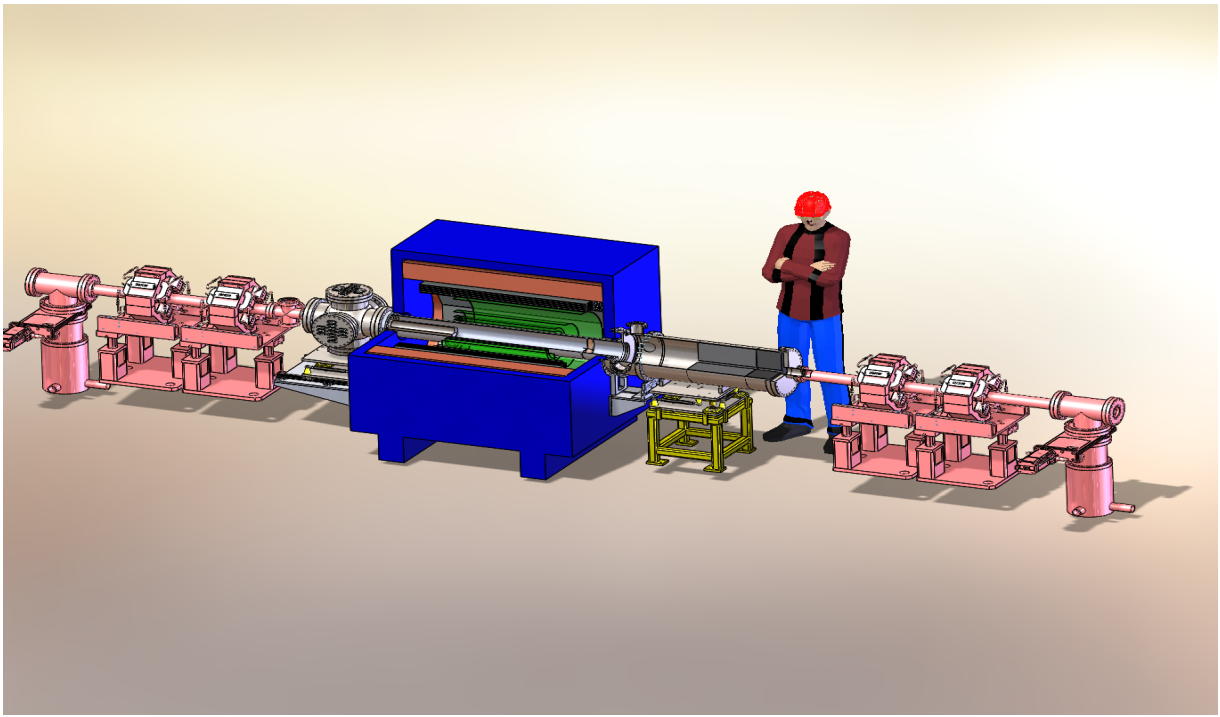}
\caption{Schematic layout of the DarkLight experiment with human figure for scale.}
\label{Layout}
\end{figure*}

The DarkLight experiment has been designed to search for evidence of $A^\prime$ production and decay to both $e^+e^-$ and invisible final-states in 100 MeV
electron scattering from a windowless hydrogen gas target. The incident energy is below the threshold for pion production to keep the final state simple. 
In contrast to previous searches, DarkLight searches at low incident energy
and aims to reconstruct the complete final-state in $ep \rightarrow epe^+e^-$.  With 5 mA of electron beam from the Jefferson Lab energy recovery linac incident
on a target of thickness $10^{19}$ cm$^{-2}$, and an ideal reconstructed energy and angular resolution, a 5$\sigma$ discovery limit region is shown~\cite{Thaler} in Fig.~\ref{Sensitivity}.  DarkLight also will carry out a 
sensitive search~\cite{Kah12} for $A^\prime$ decays to invisible products, e.g.$f \bar{f}$ of the dark sector.  
The DarkLight experiment is shown schematically in Fig.~\ref{Layout}.  The 100 MeV electron beam passes through the windowless,
hydrogen gas target.  Scattered electron, recoil proton and $e^+e^-$ decay products of the $A^\prime$ are detected
in tracking silicon and gas detectors which surround the beam-target interaction.  The experiment is located within a 0.5 Tesla solenoidal magnet which serves to 
1) guide the intense rate of forward M\o ller scattered electrons to a beam dump, 2) provide longitudinal magnetic field allowing for rigidity based determination of the momentum, and 3) shield the detectors from background.  An existing magnet from BNL-AGS  experiment E906 has been procured for the DarkLight experiment.  
The final-state leptons have momenta in the range 10-90 MeV/c so material thickness must be
kept to a minimum ($\leq$ 1 \% rad. len.) to minimize the effects of multiple scattering.  The design goal is 1 MeV/c$^2$ in reconstructed $e^+e^-$ invariant mass.
The physics processes in the DarkLight experiment are elastic and M\o ller scattering and their associated radiative processes, superimposed on a 
background of secondary charged particles and gammas from leptons and gammas traversing the DarkLight detector and backscattered from the beam dump.  
Detailed simulation codes based on Geant4 have been developed which have aided the design of the experiment.
 The DarkLight experiment drives new technology for the beam, gas target, and detectors.  The beam required has high power, high brightness, and is characterized with low halo.
The feasibility of DarkLight was demonstrated in a crucial test which established that the 0.4 Megawatt electron beam from the Jefferson Lab energy recovery linac 
was stably passed~\cite{Test} through a 2 mm diameter $\times$ 127 mm length aperture with beam losses of 6 ppm.   The windowless hydrogen gas target of thickness
10$^{19}$ cm$^{-2}$ must be designed to allow a beam of power $\sim$ 0.5 Megawatt with minimal background generation.  The detectors must accept true elastic rates of 
order 100 MHz and be read out so that a true rate of order kHz of $e^-pe^+e^-$ QED physics events is recorded with high efficiency. 

The DarkLight experiment demonstrates a new means to carry out electron scattering at low momentum transfer which has attracted international attention.
A workshop was held at MIT in March 2013 to explore scientific opportunities with this new technique~\cite{PEB13}.  Complementary DarkLight-type experiments are under
realization at Mainz, Germany~\cite{PEB13} and under consideration at Cornell University.  

\section{Path to Realization}
The DarkLight experiment received full scientific approval for 90 days running with an {\bf A} rating from Jefferson Lab in March 2013.
The DarkLight collaboration has pursued a path to realization that uses existing equipment, where feasible, and has a staged approach.
The target is under construction using elements of the windowless gas target used in the OLYMPUS experiment at DESY, Hamburg, Germany~\cite{Ber14b}.
The solenoidal magnet has been provided by the Stony Brook University group.

\subsection{Phase-I}
In 2014 a phase-I DarkLight experiment was funded by the NSF MRI program.  This would allow realization of an experiment at the Jefferson Lab ERL at 100
MeV at design luminosity with the existing magnet.  Prototype GEM trackers and silicon detectors will be used.  There are three scientific goals of the phase-I experiment: 
1) beam studies of the effect of the gas target and solenoidal magnet on the ERL beam, 2) measurement of Standard Model processes in the experiment,
3) a search for the $A^\prime$.  DarkLight phase-I would be completed in 20 days of running. The phase-I experiment will be assembled and commissioned
at the MIT-Bates Research and Engineering Center before shipping to Jefferson Lab.  It is expected that the phase-I experiment will take data in the next eighteen months.  

\subsection{Phase-II}
In phase-II the complete DarkLight experiment would be realized to search for the $A^\prime$ with the sensitivity shown in Fig.~\ref{Sensitivity}.  Further, the experiment
would be optimized for sensitivity to invisible decays of the $A^{\prime}$. The detailed design of phase-II is in progress.  The aluminum target pipe would be modified to be made from beryllium.  The silicon proton
detector would likely be a full acceptance implementation of the technology used in phase-I.  The lepton tracker under consideration consists of four
concentric, cylindrical micromegas of the type developed for the CLAS12 detector.  The experimental constraints make a classical trigger design impossible.
At present, under consideration is a streaming, dead-time free, trigger-less readout system based on fast ADCs and FPGAs.  Preliminary tracking will be performed online on a CPU farm in real time.  Data written to disk is then selected based on the tracker results.  It is anticipated that the phase-II experiment could take data within two years after completion of phase-I.

\section{Requirements}
Continued funding of the research groups in the DarkLight collaboration is essential.  Development of a triggerless readout is essential for the success of the
DarkLight experiment.   Healthy operations support for Jefferson Lab is also vital.
The equipment funds to construct the DarkLight phase-II experiment are estimated at \$ 5-7 million.

This work has been supported in part by the US Department of Energy Offices of High Energy Physics and Nuclear Physics as well
as by the National Science Foundation Divisions of Elementary Particle Physics and Nuclear Physics.  We also acknowledge generous 
support from the MIT Physics Department and School of Science.

\end{document}